\documentclass[accepted=2023-09-25, a4paper]{quantumarticle}

\pdfoutput=1

\usepackage[utf8]{inputenc}
\usepackage[english]{babel}
\usepackage[T1]{fontenc}
\usepackage{bbm}
\usepackage{amsmath}
\usepackage{amssymb}
\usepackage[numbers]{natbib}
\usepackage{hyperref}
\usepackage[export]{adjustbox}
\usepackage{float}
\usepackage{placeins}

\usepackage[noabbrev,capitalize]{cleveref}
\crefname{equation}{Eq.}{Eqs.}
\crefname{figure}{Fig.}{Figs.}
\crefname{table}{Tab.}{Tabs.}
\crefname{section}{Sec.}{Secs.}
\crefname{subsection}{Subsec.}{Subsecs.}
\crefname{subsubsection}{Subsubsec.}{Subsubsecs.}

\begin{document}

\title[Cutting multi-control quantum gates with ZX calculus]{Cutting multi-control quantum gates with ZX calculus}
\author{Christian Ufrecht}
\email{christian.ufrecht@iis.fraunhofer.de}
\address{Fraunhofer IIS, Fraunhofer Institute for Integrated Circuits IIS, Division Positioning and Networks, Nuremberg, Germany}
\author{Maniraman Periyasamy}
\address{Fraunhofer IIS, Fraunhofer Institute for Integrated Circuits IIS, Division Positioning and Networks, Nuremberg, Germany}
\author{Sebastian Rietsch}
\address{Fraunhofer IIS, Fraunhofer Institute for Integrated Circuits IIS, Division Positioning and Networks, Nuremberg, Germany}
\author{Daniel D. Scherer}
\address{Fraunhofer IIS, Fraunhofer Institute for Integrated Circuits IIS, Division Positioning and Networks, Nuremberg, Germany}
\author{Axel Plinge}
\address{Fraunhofer IIS, Fraunhofer Institute for Integrated Circuits IIS, Division Positioning and Networks, Nuremberg, Germany}
\author{Christopher Mutschler}
\address{Fraunhofer IIS, Fraunhofer Institute for Integrated Circuits IIS, Division Positioning and Networks, Nuremberg, Germany}

\begin{abstract}
\noindent
Circuit cutting, the decomposition of a quantum circuit into independent partitions, has become a promising avenue towards experiments with larger quantum circuits in the noisy-intermediate scale quantum (NISQ) era.
While previous work focused on cutting qubit wires or two-qubit gates, in this work we introduce a method for cutting multi-controlled Z gates. 
We construct a decomposition and prove the upper bound $\mathcal{O}(6^{2K})$ on the associated sampling overhead, where $K$ is the number of cuts in the circuit. This bound is independent of the number of control qubits but can be further reduced to $\mathcal{O}(4.5^{2K})$ for the special case of CCZ gates. Furthermore, we evaluate our proposal on IBM hardware and experimentally show noise resilience due to the strong reduction of CNOT gates in the cut circuits.
\end{abstract}

\maketitle

\section{Introduction}
\label{Introduction}
Quantum computing \cite{Nielsen2010} in the noisy-intermediate scale quantum (NISQ) era is limited by the 
strong impact of noise and the small number of available qubits \cite{Preskill2018}. As a result, current hardware is far from being able to execute quantum algorithms with provable quantum advantage, such as Shor's \cite{Shor1994} or Grover's \cite{Grover1996} algorithm.
To overcome these hardware restrictions to some extent, circuit-cutting techniques have recently attracted a lot of attention. When scaling up problem instances for the search of empirical quantum advantage in quantum-machine learning tasks or for the quantum approximate optimization algorithm (QAOA) \cite{Farhi2014}, these methods are expected to become relevant tools and will likely be an inherent part of near-term quantum software frameworks \cite{IBM_Roadmap_2022}. 

Consider a quantum circuit consisting of two partitions, only connected by a few wires or two-qubit gates.
By decomposing the identity channel, Peng et al.\ \cite{Peng2020} introduced \textit{wire cutting} where a qubit line is cut along the direction of time.
This method was subsequently further investigated \cite{Perlin2021, Uchehara2022, Chen2022, Ying2022} with respect to the interplay between circuit cutting and noise \cite{Ayral2020, Ayral2021, Liu2022,  Majumdar2022}, 
automatic allocation of classical and quantum computational resources \cite{Tang2021, Tang2022} and compilation \cite{Chatterjee2022}.
Similarly, Mitarai and Fujii \cite{Mitarai2019, Mitarai2021, Mitarai2021b} proposed \textit{gate cutting}, the direct decomposition of a unitary channel corresponding to a two-qubit gate.
As in probabilistic error mitigation \cite{Temme_2017,Piveteau_2022_quasiprob, Piveteau_2020,Mitarai2021b} the variance of the estimator for the quantity to be measured increases \cite{Pashayan_2015} in case of a cut circuit.
As a consequence, all variants of circuit cutting come with a constant $\kappa$, characterizing the sampling overhead, the factor $\mathcal{O}(\kappa^2)$ of more samples required to estimate the decomposed circuit to the same accuracy as the original one. If $K$ cuts are performed, the sampling overhead increases to $\mathcal{O}(\kappa^{2K})$.
Of course, the exponential overhead is in line with
our expectation of classical hardness for simulation of general quantum circuits.
Piveteau et al.\ \cite{Piveteau2022_circuitcut} showed that this overhead can be significantly reduced when a common decomposition of several Bell states with subsequent gate teleportation is performed. Ref.\ \cite{Piveteau2022_circuitcut} also provides optimal decompositions for several important two-qubit gate types  based on the robustness of entanglement measure \cite{Vidal_1999}. A strong reduction of sampling overhead can also be achieved for joint cutting of wires as shown by Lowe et al.\ \cite{Lowe2022}.
\begin{figure}
    \centering
    \includegraphics[width=7.8cm]{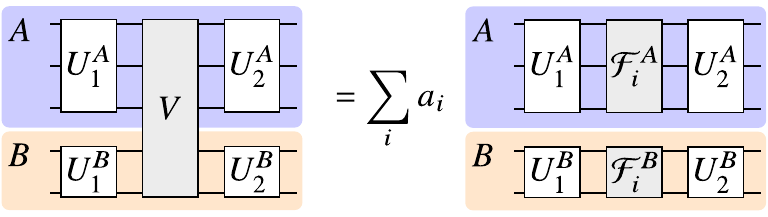}
    \caption{Consider the quantum circuit shown on the left.
    It is a circuit with five qubits represented by the horizontal lines, also referred to as wires. Time evolution of the initial state is illustrated by the white and grey boxes where time flows from the left to the right. They represent quantum gates, that is, unitary transformations on subsets of the qubits.
    The circuit contains a single gate $V$ that connects the two partitions $A$ and $B$ that are otherwise independent. After decomposition of the quantum channel corresponding to $V$, the circuit disintegrates into a weighted sum over independent circuit pairs, which contain quantum gates or measurements denoted by $\mathcal{F}_i^A$ and $\mathcal{F}_i^B$. Here, we employ the calligraphic notation otherwise reserved for superoperators to indicate that the gate or projectors are determined from the superoperator decomposition.
    The result of the original circuit can be restored by evaluating sequentially each of the circuits on possibly smaller quantum devices. 
    Note that the equality has to be understood on the superoperator rather than on the gate level.}
    \label{quantum_circuit}
\end{figure}

\begin{table}
\caption{Overview of recent findings on CZ and wire cutting. We contrast the sampling overhead for $K$ cuts associated with different methods and compare them to the main results of the present work.}
\label{decomposition of different gate types}
\centering
\begin{tabular}{ p{1.7cm}p{3.2cm}p{1.9cm} }
             &  sampling overhead  & joint cutting \\
  \hline
  \vspace{0.1pt}
   CZ gate   &    \vspace{0.1pt}$\mathcal{O}(3^{2K})$ \cite{Mitarai2021}  & \vspace{0.1pt} $\mathcal{O}(4^K)$ \cite{Piveteau2022_circuitcut} \\
   \vspace{0.1pt}
   wire &  \vspace{0.1pt}$\mathcal{O}(4^{2K})$ \cite{Peng2020} & \vspace{0.1pt}$\mathcal{O}(4^K)$ \cite{Lowe2022}\\
   \vspace{0.1pt}
   CCZ gate & \vspace{0.1pt}$\mathcal{O}(4.5^{2K})\;$ [this work] & \vspace{0.1pt}$\quad-$\\
    \vspace{0.1pt}
   MCZ gate & \vspace{0.1pt}$\mathcal{O}(6^{2K})\;$ [this work] & \vspace{0.1pt}$\quad-$\\
\end{tabular}
\end{table}
In this work, we provide explicit decompositions of multi-controlled $Z$ (MCZ) gates with the help of ZX-calculus, a tensor-network description for quantum circuits. 
After explaining the ideas underlying gate-cutting in more detail and comparing the sampling overhead of different methods in \cref{Spatial cuts}, we prove the main result of this article, the general decomposition of an MCZ gate in \cref{ZX-calculuas for circuit cutting} with the help of ZX calculus. In \cref{Discussion}, we evaluate the sampling complexity associated with the decomposition of an MCZ gate.
We show the upper bound $\kappa=6$, which, remarkably, is independent of the number of control qubits. For a CCZ gate, we find the smaller value $\kappa=4.5$. We conclude this article with experimental results on the IBM Q system Ehningen discussed in \cref{Experiments on IBMQ}.
We observe a strong reduction of the CNOT-gate count in the cut circuits and, therefore, resilience to noise.
As a particular application of MCZ-gate cutting, we anticipate its use in the alternating operator ansatz \cite{Saleem2021, Hadfield2018}, a variant of the QAOA \cite{Farhi2014} algorithm for constrained optimization. Another promising application is the simulation of MCZ gates connecting qubits far apart on the hardware graph as done for two-qubit gates in Refs.\ \cite{Mitarai2021,Yamamoto2022}.

\section{Cutting a quantum gate}
\label{Spatial cuts}
Consider a quantum circuit where the qubits are grouped into two partitions $A$ and $B$, only connected by one possibly multi-qubit gate $V$ as shown in \cref{quantum_circuit}. Furthermore, assume the factorizing initial state $\rho=\rho^A\otimes \rho^B$ and observable $O=O^A\otimes O^B$ where the superscript labels the partition. These assumptions are, for example, satisfied when all qubits are initialized to $|0\rangle$ and a Pauli string is measured.
Circuit cutting, also referred to as circuit decomposition, circuit fragmentation or circuit knitting, is the task of finding a decomposition of the unitary channel $\mathcal{V}$ corresponding to the gate $V$ so that
\begin{equation}
\label{general_decomposition_formula}
    \mathcal{V}=\sum_i a_i \mathcal{F}_i
\end{equation}
where $\mathcal{F}_i(\rho^A\otimes \rho^B)=\mathcal{F}_i^A(\rho^A)\otimes \mathcal{F}_i^B(\rho^B)$ are local channels on the partitions $A$ and $B$ as shown in \cref{quantum_circuit} and $a_i$ are real expansion coefficients.
For $\mathcal{U}_1$ corresponding to $U_1^A \otimes U_1^B$ and $\mathcal{U}_2$ corresponding to $U_2^A \otimes U_2^B$, this decomposition allows us to determine the expectation value of the observable as 
\begin{align}
    \label{Definition_cutting1}
    \langle O\rangle
    &=\mathrm{tr}(O \mathcal{U}_2
    \circ \mathcal{V} \circ \mathcal{U}_1(\rho))\\
    \label{Definition_cutting2}
    &=\sum_i a_i \mathrm{tr}(O^A\otimes O^B \mathcal{U}_2 \circ \mathcal{F}_i \circ \mathcal{U}_1(\rho^A\otimes\rho^B))\\
    \label{Definition_cutting3}
    &=\sum_i a_i \langle O^A\rangle_i \langle O^B\rangle_i\,.
\end{align}
In \cref{Definition_cutting3} we denote by $\langle.\rangle_i$ the expectation value with respect to the state on partition $A$ and $B$, evolved by $\mathcal{F}_i^A$ and $\mathcal{F}_i^B$, respectively, that is
\begin{equation}
\langle O^\alpha\rangle_i=\mathrm{tr} (O^\alpha \mathcal{U}_2^\alpha \circ \mathcal{F}_i^\alpha \circ \mathcal{U}_1^\alpha(\rho^\alpha))
\end{equation}
for $\alpha=A,B$.
 If $\mathcal{F}_i^\alpha$ are unitary channels or measurement operations themselves, each term in the sum in \cref{Definition_cutting3} can be evaluated on a quantum computer.
In the case of a more general quantum circuit, when all gates connecting the two partitions are cut, they become independent and the quantum circuits can be evaluated sequentially on a smaller device.
We emphasize that the term circuit cutting as used here deals with decompositions of quantum circuits at the level of superoperators, rather than at the level of unitaries as in Ref.\ \cite{Marshall2022}.
The output of a single experimental shot on a quantum computer is typically a bitstring. The expectation value of an observable is then obtained by a classical post-processing function on the bitstrings of multiple runs. Modeling the outcome of each experimental run as i.i.d.\ random variables, the number of required samples to achieve a given additive error is determined via the variance of the post-processing function.
When a circuit is cut, the variance of the modified estimator for $\langle O\rangle$ via \cref{Definition_cutting3} increases.
Consequently, more experimental runs are required to estimate the result of the original circuit to the same given additive error. More specifically, this sampling overhead is exponential (in the number of cuts).
The parameter
\begin{equation}
\kappa=\sum_i |a_i|
\label{definition_kappa}
\end{equation}
then quantifies the sampling overhead $\mathcal{O}(\kappa^2)$ \cite{Pashayan_2015}. 
For completeness, this scaling behavior is re-derived in \cref{Sampling-complexity overhead}.
\cref{general_decomposition_formula} is therefore optimal if the   $\mathcal{F}_i$ are chosen such that the 1-norm of the vector containing the coefficients $a_i$ is minimal.
For $K$ cuts the sampling overhead is $\mathcal{O}({\kappa^{2K}})$, however, joint cutting of multiple gates or wires leads to much smaller bounds for some gate types \cite{Piveteau2022_circuitcut, Lowe2022}. \cref{decomposition of different gate types} summarizes recent findings for cutting of CZ gates and wire cutting and compares the sampling complexities to the main results of the present work.
In this work we cut multi-qubit controlled Z gates and provide upper bounds for $\kappa$.
As we will explore in the next section, ZX calculus is particularly suited for this task.

\section{ZX-calculus for circuit cutting}
\label{ZX-calculuas for circuit cutting}
ZX calculus \cite{Coecke_2011, Coecke_2008} is a tensor-network representation for quantum circuits that together with powerful transformation rules allows diagrammatic reasoning. Since ZX-calculus has been reviewed elsewhere \cite{Wetering2020}, we will only introduce those diagram types and transformation rules necessary for this article.
The basic diagrams are Z-spiders, defined as
\begin{align}
\label{Z_spider}
{\scriptstyle m}\;\vcenter{\hbox{\includegraphics[width=0.9cm]{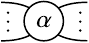}}}\;{\scriptstyle n}
\;\;&= \underbrace{|0...0\rangle}_{\scriptstyle n} \underbrace{\langle 0...0|}_{\scriptstyle m} +\mathrm{e}^{i\alpha }|1...1\rangle \langle 1...1|\\
\intertext{and X-spiders}
\label{X_spider}
{\scriptstyle m}\;\vcenter{\hbox{\includegraphics[width=0.9cm]{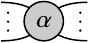}}}\;{\scriptstyle n}
\;\;&=\underbrace{|{\scriptstyle+...+}\rangle}_{\scriptstyle n}\underbrace{\langle {\scriptstyle+...+}|}_{\scriptstyle m}+\mathrm{e}^{i\alpha}|{\scriptstyle-...-}\rangle\langle {\scriptstyle-...-}|\\
\intertext{ where $|\pm\rangle=(|0\rangle \pm |1\rangle)/\sqrt{2}$. For $\alpha=0$, the inset is commonly disregarded. A third tensor, a so-called H-box, is defined as} 
\label{Hbox}
{\scriptstyle m}\;\vcenter{\hbox{\includegraphics[width=0.95cm]{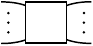}}}\;{\scriptstyle n}
\;\;&=\sum(-1)^{i_1\cdot...\cdot i_m\cdot j_1\cdot...\cdot j_n}|j_1...j_n\rangle\langle i_1...i_m|
\end{align}
where the sum runs over all $i_1,...,i_m, j_1,...,j_n\in \{0,1\}$.
Spiders and H-boxes are therefore maps from $m$- to (un-normalized) $n$-qubit states, signified by the number of wires ending at the right and the left of the diagrams.
When representing ZX diagrams as matrices, we will implicitly assume the computational basis. Then, for example, an H-box is a matrix filled with ones but a minus one in the lower right corner.  H-boxes can be viewed as generalized Hadamard gates since
\begin{equation}
\label{Hadamard}
 \vcenter{\hbox{\includegraphics[width=0.5cm]{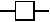}}}=\sqrt{2}\,
  \vcenter{\hbox{\includegraphics[width=0.84cm]{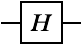}}}\,.
\end{equation}
As apparent from the definitions, spiders and H-boxes correspond to symmetric tensors in all indices. Any quantum circuit can be represented as a ZX diagram, the reverse statement, however,  is incorrect since a ZX diagram not necessarily corresponds to a unitary matrix. To obtain the Hermitian conjugate of a ZX diagram, we move all wires ending at the left to the right and those ending at the right to the left and replace all angles with their negative values. In the following we show how to use ZX calculus to decompose the unitary channel corresponding to an MCZ gate into a sum over unitary and measurement channels. Decomposing circuits into simpler parts using ZX calculus has been done before in Refs. \cite{Kissinger_2022a, kissinger_2022b}.

An MCZ gate enjoys a simple representation \cite{Backens_2019} 
\begin{equation}
\label{CZ-gate}
 \vcenter{\hbox{\includegraphics[width=0.56cm]{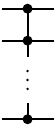}}}\quad=\quad
  \vcenter{\hbox{\includegraphics[width=1.42cm]{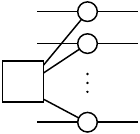}}}
\end{equation}
in terms of an H-box with zero wires on the left as shown in \cref{Representation of a MCZ gate}. In matrix representation, $MCZ=\mathrm{diag}(1,...,1,-1)$ and we will refer to the number of qubits involved in the MCZ gate as the \textit{order} of the gate. Also note that there is no difference between the control and target qubits in an MCZ gate. The decomposition constructed in this article is based on the H-box fusion rule \cite{Backens_2019}
\begin{equation}
\includegraphics[width=1.3cm,valign=c]{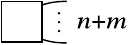}\; = \frac{1}{2}\;\includegraphics[width=1.37cm,valign=c]{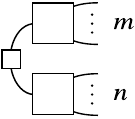}\,,
 \label{fusion_rule}
\end{equation}
proved in \cref{Proof of H-box fusion rule} for completeness.

Consequently, with \cref{fusion_rule} applied to \cref{CZ-gate}, the unitary channel action $\mathcal{E}_{MCZ}(\rho)=MCZ \rho MCZ^\dagger$ corresponding to the MCZ gate of order $n+m$ acting on an arbitrary density matrix $\rho$, takes the form
\vspace{1.3cm}
\begin{equation}
\label{Hbox_decomposition}
 \mathcal{E}_{MCZ}(\rho)=\frac{1}{4}\;\,\vcenter{\vspace{-1.3cm}\hbox{\includegraphics[width=3.77cm]{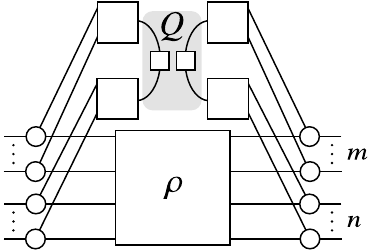}}}\,.
\end{equation}
The two partitions (upper $m$ qubits, lower $n$ qubits) are only connected by the tensor $Q$ shaded in grey, the un-normalized Choi operator of a Hadamard gate.
Next, we remove the two remaining connections between the partitions by a rank-one decomposition of $Q$ in terms of vectors factorizing over the two partitions
\begin{align}
Q&=
\vcenter{\hbox{\includegraphics[width=0.80cm]{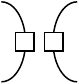}}}
=\begin{pmatrix}
 1 & 1 & 1 &-1\\
1 & 1 & 1 &-1\\
1 & 1 & 1 &-1\\
-1 & -1 & -1 &1
\end{pmatrix}\\
&=\sum_i c_i
\vcenter{\hbox{\includegraphics[width=1.74cm]{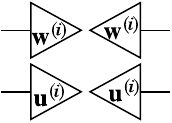}}}\,.
\label{rank_one_decomp}
\end{align}
 In \cref{rank_one_decomp} we introduced the diagrammatic notation for a two-dimensional vector
\begin{equation}
  \mathbf{v}=\sum_{j\in \{0,1\}} v_j |j\rangle =\; \vcenter{\hbox{\includegraphics[width=0.86cm]{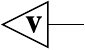}}}
   \;=\;\left( \vcenter{\hbox{\includegraphics[width=0.86cm]{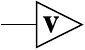}}}\right)^\dagger
\end{equation}
where $v_j$ is the $j$-th  component of $\mathbf{v}$.
Therefore, the H-box fusion rule in \cref{Hbox_decomposition} reduces the cutting of an MCZ gate to the decomposition of a four-by-four matrix.
However, only vectors $\mathbf{w}^{(i)}$ and $\mathbf{u}^{(i)}$ are useful whose contraction with the H-box results in channels that can be evaluated on a quantum computer. In \cref{H-box identities} we prove the identities 
\begin{align}
\label{main_identity1}
    \vcenter{\hbox{\includegraphics[width=2.06cm]{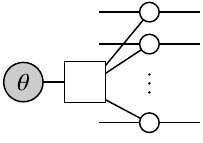}}}\quad&=\sqrt{2}\;\;\vcenter{\hbox{\includegraphics[width=0.86cm]{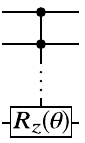}}}\\
    \intertext{for $\theta \in [0,2\pi)$ and}
    \label{main_identity2}
    \vcenter{\hbox{\includegraphics[width=2.06cm]{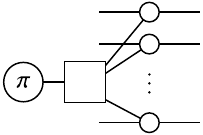}}}\quad&=2\hspace{0.35cm}\vcenter{\hbox{\includegraphics[width=1.04cm]{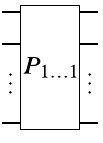}}}
\end{align}
where $P_{1...1}$ denotes the projector on $|1...1\rangle$.
The identities \cref{main_identity1} and \cref{main_identity2} suggest 
$\mathbf{w}^{(i)}, \mathbf{u}^{(i)} \in \left\{\vcenter{\hbox{\includegraphics[width=1.4cm]{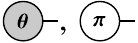}}}\right\}_\theta$
for different values of $\theta$. 
This choice guarantees a decomposition consisting of unitary controlled Z-rotation gates and projectors, which can be evaluated by mid-circuit measurements. 
 Next, we expand $Q$ in the basis spanned by the Pauli matrices and then substitute their spectral representation
\begin{align}
\label{Pauli_X}
\vcenter{\hbox{\includegraphics[width=4.2cm]{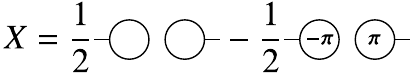}}}\\
\label{Pauli_Y}
\vcenter{\hbox{\includegraphics[width=4.2cm]{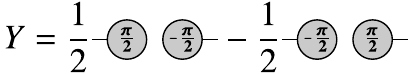}}}\\
\label{Pauli_Z}
\vcenter{\hbox{\includegraphics[width=4.2cm]{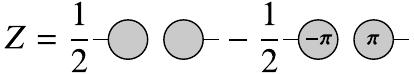}}}
\end{align}
and additionally the identity matrix in terms of the eigenvectors of the Pauli $Y$ matrix, that is 
$\mathbb{I}=
1/2\,\vcenter{\hbox{\includegraphics[width=1.32cm]{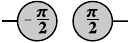}}}
+1/2\,\vcenter{\hbox{\includegraphics[width=1.32cm]{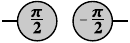}}}$. The choice of this expansion basis leads to $\kappa=3$ for the normalized version of $Q$ which can be shown to be optimal with the help of the robustness of entanglement measure  \cite{Piveteau2022_circuitcut, Vidal_1999}.
Unfortunately, \cref{Pauli_X} contains the state $\vcenter{\hbox{\includegraphics[width=0.4cm]{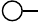}}}$, not covered by \cref{main_identity1} and \cref{main_identity2} and its contraction with an H-box does not result in a valid quantum circuit.
We could simulate this operation by an MCZ gate on an ancilla qubit initialized in and postselected on the $|+\rangle$ state \cite{Wetering}, however, at the cost of one ancilla qubit per partition and cut.
We circumvent this issue by substituting $X=1/2\,\vcenter{\hbox{\includegraphics[width=0.9cm]{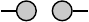}}}+1/2\,\vcenter{\hbox{\includegraphics[width=1.32cm]{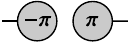}}}-\vcenter{\hbox{\includegraphics[width=1.32cm]{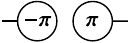}}}$ instead.
By this substitution we lose optimality for the decomposition of $Q$. But note that it is unknown if an optimal decomposition of $Q$ translates into an optimal decomposition of the MCZ gate. While we show only upper bounds, we observe in \cref{Discussion} the reduction of the general decomposition of an MCZ gate for a CZ gate to the optimal one. With these considerations in mind and defining $\Theta=\{-\pi/2, 0, \pi/2, \pi\}$, we find
\begin{align}
    Q&=\mathbb{I}\otimes \mathbb{I}+Y\otimes Y+Z\otimes X+X\otimes Z\\
    \nonumber
&=\frac{1}{2}\sum_{\theta\in \Theta}\alpha_\theta
\vcenter{\hbox{\includegraphics[width=1.3cm]{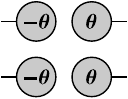}}}\\
\label{decomposition of Q}
&-
%-----------------
\frac{1}{2}\sum_{\theta\in \{0, \pi\}}\alpha_\theta\left\{
\vcenter{\hbox{\includegraphics[width=1.3cm]{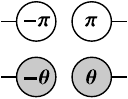}}}
+
%--------
\vcenter{\hbox{\includegraphics[width=1.3cm]{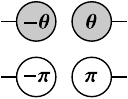}}}
\right\}
\end{align} 
where $\alpha_\theta=-1$ for $\theta=\pi$ and $\alpha_\theta=1$ otherwise. 
This result leads to a general decomposition of the MCZ gate by substituting \cref{decomposition of Q} into \cref{Hbox_decomposition} and then contracting the vectors with the H-boxes with the help of \cref{main_identity1} and \cref{main_identity2}.
Before we state the resulting decomposition explicitly, we rewrite the superoperator $\mathcal{P}_{1...1}$ corresponding to the projector $P_{1...1}$ in \cref{Rewriting the Projector} with the result
\begin{equation}
\label{projector1}
    2\mathcal{P}_{1...1}=\mathcal{Z}\,-\,\mathcal{P}\,.
\end{equation}
Here, the operation $\mathcal{Z}$ abbreviates the sum over the unitary channels corresponding to all combinations of one-qubit identities and Z gates, that is for an $n$-qubit state $\rho$
\begin{equation}
\label{Definition_ZE}
    \mathcal{Z}(\rho)=\frac{1}{2^n}\sum_{k \in \{0,1\}^n}Z^{k_1}\otimes...\otimes Z^{k_n}\rho  Z^{k_1}\otimes...\otimes Z^{k_n}\,.
\end{equation}
The second operation is  $\mathcal{P}(\rho)=\sum_l \xi_l P_l \rho P_l$ where $P_l$ are the projectors on all elements of the computational basis and $\xi_l=-1$ for $l=(1,...,1)$ as well as $\xi_l=1$ otherwise. 
We show in \cref{Sampling-complexity overhead} how to relate $\mathcal{P}$ to circuits with intermediate measurements. Furthermore, we prove that circuits containing either $\mathcal{P}$ or $\mathcal{P}_{1...1}$ have the same sampling complexity, but introducing $\mathcal{Z}$ leads to gate cancellations in the final decomposition. The consequence is a reduction of $\kappa$ in the final result. 

We now state the main result of this article, the decomposition of an MCZ gate:
\begin{widetext}
\begin{equation}
\vcenter{\hbox{\includegraphics[width=0.68cm]{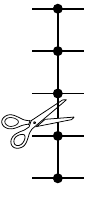}}}\;=\;
\frac{1}{2}\;\vcenter{\hbox{\includegraphics[width=0.448cm]{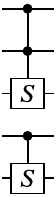}}}
\quad+\frac{1}{2}\;\vcenter{\hbox{\includegraphics[width=0.448cm]{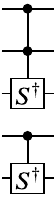}}}
\quad+\frac{1}{2}\sum_{j\in\{0,1\}}(-1)^j\left\{\;\vcenter{\hbox{\includegraphics[width=0.448cm]{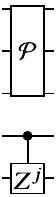}}}
\quad+\quad\vcenter{\hbox{\includegraphics[width=0.448cm]{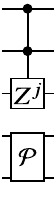}}}\;\right\}
\quad+\frac{1}{2}\sum_{j\in\{0,1\}}(-1)^j\left\{\;\vcenter{\hbox{\includegraphics[width=0.448cm]{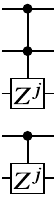}}}
\;-\;\vcenter{\hbox{\includegraphics[width=0.448cm]{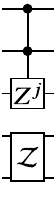}}}
\;-\;\vcenter{\hbox{\includegraphics[width=0.448cm]{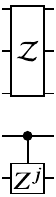}}}\;\right\}
\label{main_result}
\end{equation}
\end{widetext}
Here, $\mathrm{S}=R_z(\pi/2)$  denotes an S gate and
\begin{equation}
 \vcenter{\hbox{\includegraphics[width=1.2cm]{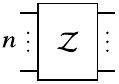}}} \quad=\quad \frac{1}{2^n}\sum_{i\in\{0,1\}^n}\;
 \vcenter{\hbox{\includegraphics[width=1.0cm]{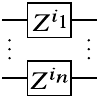}}}
\label{projector2}
\end{equation}
symbolizes the operation $\mathcal{Z}$ in circuit form.
The decomposition shown here is for an MCZ gate of order five and a cut after the third qubit line counting from the top.
We emphasize that since the derivation was independent of the order of the MCZ gate and the cutting location, the structure of the decomposition in \cref{main_result} remains unchanged for a cut at any position and arbitrary order of the MCZ gate.
\cref{main_result} has to be read in terms of channels. To evaluate a quantum circuit containing an MCZ gate we have to subsequently replace the gate by the gates and operations shown in \cref{main_result}.

\section{Sampling overhead}
\label{Discussion}
In this section we investigate the sampling overhead associated with \cref{main_result} for different orders of the MCZ gate. The sampling overhead $\mathcal{O}(\kappa^2)$ is characterized by $\kappa=\sum_i |a_i|$ as stated in \cref{definition_kappa}. We re-derive this statement in detail in \cref{Sampling-complexity overhead}. We also consider operations such as $\mathcal{P}$ and show that with regard to sampling overhead, this operation can be treated in the same manner as a unitary channel. With this insight, we are now in the position to determine $\kappa$ for \cref{main_result}. Recalling the pre-factor $1/2^n$ for the $2^n$ unitary circuits summarized by $\mathcal{Z}$ in \cref{projector2}, we find for a general MCZ gate $\kappa=6$. Remarkably, this bound holds independently of the order of the gate.
In the right bracket of \cref{main_result} the MCZ$^j$ gate is the identity gate for $j=0$ and $\mathcal{Z}$ also contains the identity gate but the prefactors have a different sign.
Consequently, summing these gates slightly reduces $\kappa$. This reduction becomes more pronounced, the smaller the order of the MCZ gate is but quickly approaches $\kappa=6$ for large order. Indeed, for a CZ gate, the right bracket vanishes and we find $\kappa=3$ which is known to be optimal \cite{Piveteau2022_circuitcut}. Similarly, a cut separating one qubit line from a general MCZ gate results in $\kappa=5$. For a CCZ gate, we find $\kappa=4.5$. Due to the importance of CCZ gates in quantum computing theory, its decomposition is stated explicitly:
\begin{widetext}
\begin{equation}
\vcenter{\hbox{\includegraphics[width=0.86cm]{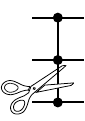}}}\;=\;
\frac{1}{2}\;\vcenter{\hbox{\includegraphics[width=0.56cm]{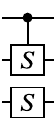}}}
\;+\;\frac{1}{2}\;\vcenter{\hbox{\includegraphics[width=0.56cm]{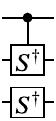}}}
\;+\;\frac{1}{2}\sum_{j\in\{0,1\}}(-1)^j\left\{\;\,\vcenter{\hbox{\includegraphics[width=0.56cm]{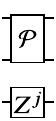}}}
\;+\;\vcenter{\hbox{\includegraphics[width=0.56cm]{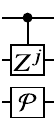}}}\;\right\}
\;+\;\frac{1}{4}\sum_{j\in\{0,1\}}(-1)^j\left\{\;\vcenter{\hbox{\includegraphics[width=0.56cm]{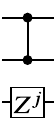}}}
\;\,-\;\,2\;\vcenter{\hbox{\includegraphics[width=0.56cm]{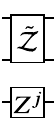}}}\;\right\}
\label{CCZgate}
\end{equation}
\end{widetext}
with the abbreviation
\begin{equation}
\vcenter{\hbox{\includegraphics[width=0.56cm]{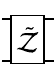}}}\;=\;\frac{1}{4}\left\{
\vcenter{\hbox{\includegraphics[width=0.56cm]{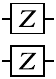}}}\;\;+\;\;
\vcenter{\hbox{\includegraphics[width=0.56cm]{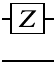}}}\;\;+\;\;
\vcenter{\hbox{\includegraphics[width=0.56cm]{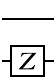}}}\;\;-\;\;
\vcenter{\hbox{\includegraphics[width=0.56cm]{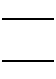}}}\right\}\,.
\end{equation}
The sampling complexities derived above are summarized in \cref{overhead_for_different_orders} together with known results on the optimality of the decomposition.
\begin{table}[httb]
\caption{Sampling overhead for different order and cutting positions for an MCZ gate. In addition, we state known results on the optimality of the decomposition.}
\label{overhead_for_different_orders}
\centering
\begin{tabular}{ p{3cm}p{1.75cm}p{1.75cm} }
   gate type & $\kappa$  & optimality  \\
  \hline
   CZ gate   &    3      & optimal \cite{Piveteau2022_circuitcut}\\
   CCZ gate  &    4.5    & unknown\\
   one qubit removed & 5 & unknown\\
   general       & 6    & unknown\\
\end{tabular}
\end{table}

\section{Experiments on IBM Q}
\label{Experiments on IBMQ}
In this section, we show experimental results obtained on the ibmq\_ehningen system.
We will find a strong reduction of noise impact for the proposed cutting scheme.

We first run numerical simulations to validate the proposed MCZ cutting scheme.
To this end, we generate a set of random 3, 4 and 5 qubit circuits with one MCZ gate at the center of the circuit and two otherwise independent partitions. For exemplary circuits and a more detailed description of the circuit generation, see Supplemental Material. 
In all experiments, we measure the $Z\otimes...\otimes Z$ Pauli string. For the cut circuits, we allocate $N_i=N|a_i|/(2\kappa)$ samples to each circuit pair labeled by $i$ where $N= 4\kappa^2/\epsilon^2$ bounds the standard deviation by $\epsilon$, see \cref{error_bound_eq} in \cref{Sample complexity by pre-estimating circuits}.
For an uncut circuit, we allocate all $N$ samples for evaluation. 
We first calculate $N=1.44 \times 10^6$ and $N=1.44 \times 10^8$ via \cref{error_bound_eq}, the number of repetitions required to bound the standard deviation of expectation values of the cut circuit to within $\epsilon=0.01$ and $\epsilon=0.001$. 
For fixed number of qubits we generate 5 random circuits and sample each circuit $N$ times using an ideal simulator. Repeating this process 20 times,  generates one hundred data points for each value of $\epsilon$. 

We now compare the distributions of the expectation-value differences  between the full (that is uncut) and cut circuits for different values of $\epsilon$.
\begin{figure}[hb]
    \centering
    \includegraphics[width=\linewidth]{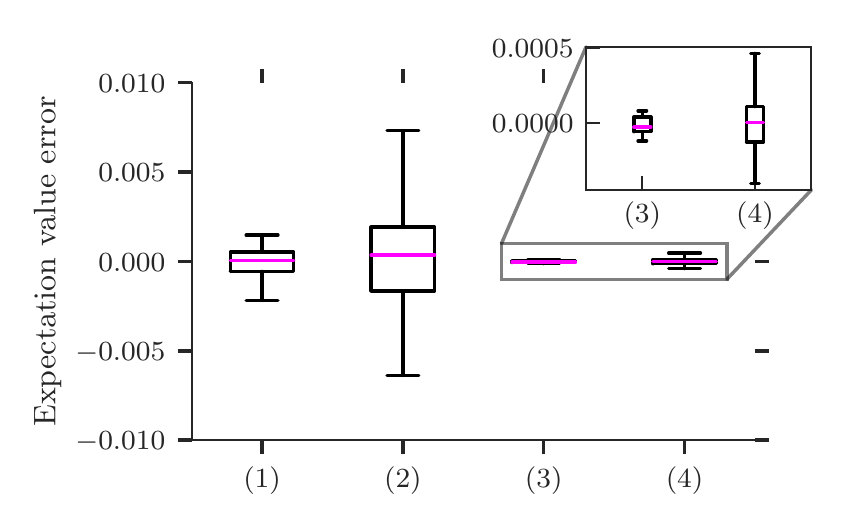}
    \caption{Expectation value estimation error for random circuits with 5 qubits, containing an MCZ gate of order 5 in their center: (1) Full circuit sampled for $1.44 \times 10^6$ repetitions, (2) Proposed method sampled for $1.44 \times 10^6$ repetitions. As expected, the distribution of results becomes broader for the cut circuits. 
    (3) Full circuit sampled for $1.44 \times 10^8$ repetitions, (4) Proposed method sampled for $1.44 \times 10^8$ repetitions. In accordance with \cref{error_bound_eq}, a factor of hundred more samples increases the accuracy by approximately a factor of ten. The values of the standard deviations in the figure are: (1): $8.3\times 10^{-4}$, (2):  $2.7\times 10^{-3}$, (3) 
 $8.4\times 10^{-5}$, (4): $2.4\times 10^{-4}$. }
    \label{shot_error_plot}
\end{figure}
We run simulations for 3,4 and 5 qubits and find almost no difference in the standard deviations of the distributions. This behavior is expected since $\kappa$ is independent from the order of the MCZ gate.
Exemplarily, \cref{shot_error_plot} shows the distribution for random 5-qubit circuits.

In the figure, the white boxes are the interquartile ranges and the bars denote the 5\% and 95\% quantiles.
\cref{shot_error_plot}(1) shows the distribution of the expectation values for the original uncut circuits. As expected, the distribution becomes broader when we instead evaluate the cut circuit in 
\cref{shot_error_plot}(2). Its standard deviation is $0.0027$,  which is smaller than the chosen value of $\epsilon=0.01$.  The same effect is observed  for $\epsilon=0.001$ shown in the plots in \cref{shot_error_plot}(3) and \cref{shot_error_plot}(4). Consequently, a tighter version of \cref{error_bound_eq} to bound the variance might exist. 
\begin{figure}
    \centering
    \includegraphics[width=\linewidth]{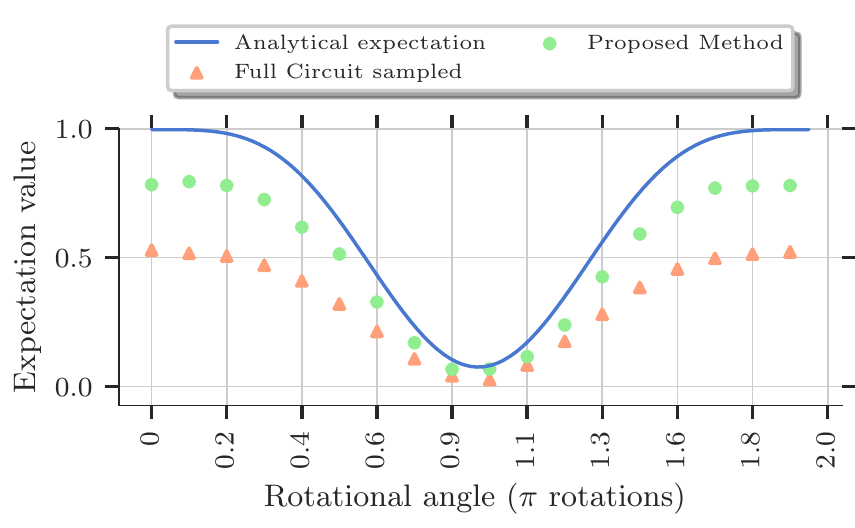}
    \caption{Expectation value estimation of a parameterized circuit containing a 5-qubit MCZ gate on a noisy simulator. Comparison to \cref{exp_5MC} shows that the noisy simulation strongly underestimates the influence of noise for sampling the uncut circuit (orange triangles). For the simulations of noise, qiskit's noise model from ibmq\_ehningen is used. The noise model is updated based on current calibration and error data in the cloud.     
    }  
    \label{exp_5MC_noisy}
\end{figure}
To further validate the proposed method on real quantum hardware and analyze the impact of noise, we performed MCZ gate-cutting experiments on the ibmq\_ehningen \cite{ibmq2021} device. The ibmq\_ehningen device is a super-conducting qubit type quantum hardware with a 27 qubit ibmq\_falcon processor. The native gate set on this hardware consists of  Rotational-Z, Pauli-X, $\sqrt{\mathrm{X}}$, Identity, and CNOT gates. For the experiments, we again generated a random circuit with one free parameter that was scanned in the experiments.
\begin{figure}
    \centering
    \includegraphics[width=\linewidth]{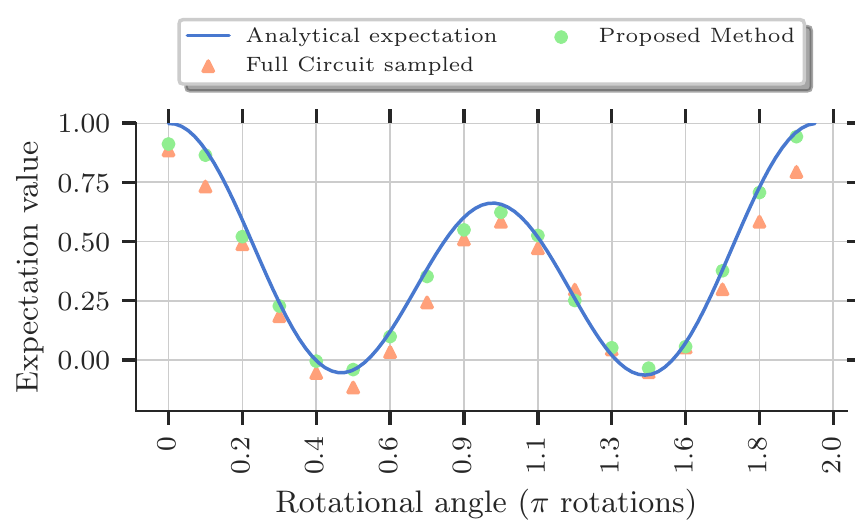}
    \caption{Expectation value estimation of parameterized circuit containing a CCZ gate on ibmq\_ehningen device. The green circles, sampled from the cut circuits, seem to lie closer to the exact result (blue line) than the result sampled from the uncut circuit (orange triangles). However, more simulations would be needed to verify that cutting a CCZ gate leads to noise reduction for the circuit considered. The x-axis shows the values of the free parameter in the circuit, see Supplementary Material.}
    \label{exp_taffoli}
\end{figure}
The maximum number of shots per job on the ibmq\_ehningen device is $10^5$ and the same was used for each data point, see Supplemental Material for the explicit circuits and a more detailed discussion of the experimental setup.
\cref{exp_taffoli} shows the result for a three-qubit circuit containing a CCZ gate. In the figure, the results obtained by cutting the CCZ gate (green circles) lie slightly closer to the exact curve (blue line) than the results from the uncut circuit (orange triangles). 

\begin{figure}
    \centering
    \includegraphics[width=\linewidth]{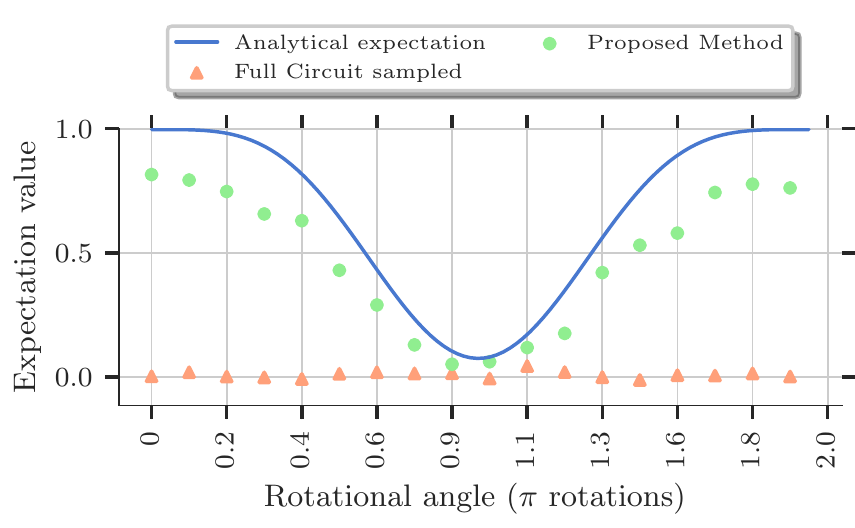}
    \caption{Expectation value estimation of parameterized circuit containing a 5-qubit MCZ gate on the ibmq\_ehningen device. The green circles, sampled from the cut circuits, still qualitatively match the exact result (blue line). The signal sampled from the uncut circuit (orange triangles) is completely derogated due to the strong influence of noise.}
    \label{exp_5MC}
\end{figure}

Next, we repeated the experiment with a five-qubit circuit containing an MCZ gate of order five. The result is shown in \cref{exp_5MC}.
From the figure, it is clear that the influence of noise is extreme when executing the transpiled version of the full circuit with an MCZ gate of order five. The resulting expectation values were nothing but a random output centered around zero. However, the expectation value estimated using the proposed method falls much closer to the  expectation values obtained by ideal simulation. This resilience to noise can be attributed to the strong reduction of the CNOT-gate count in the cut circuits. While, asymptotically, an MCZ gate of order $n$ can be synthesized with $\mathcal{O}(n)$ CNOT gates with one auxiliary qubit \cite{Barenco_1995}, for small $n$ the number of required CNOT gates increases quickly. Consequently, cutting for example two qubits from an MCZ gate of order five reduces the maximum number of CNOT gates in a circuit to those required to synthesize a double-controlled S gate.
This reduction becomes even more pronounced after compiling to topologically restricted hardware. As a consequence, we observe a strong reduction of noise impact.
The maximum number of CNOT gates present in the different circuits are shown in \cref{num_CNOT}. The number of CNOT gates might vary based on the transpilation method used. However, we chose to proceed with the best transpiler offered by the hardware manufacturer to validate our results.
\begin{table}[h]
\caption{Number of CNOT gates present in the transpiled version of the circuits shown in the Supplementary Material (first column). After circuit cutting, the maximum number of CNOT gates contained in one of the cut circuits reduces considerably (second column). }
\label{num_CNOT}
\centering
\begin{tabular}{ p{3cm}p{1.75cm}p{1.75cm}p{1.75cm} }
   gate type &  CNOT & CNOT cut\\
  \hline
   CCZ gate  &    13 & 3   \\
   5-qubit MCZ &  114 & 32\\
\end{tabular}
\end{table}
As most of the work in the literature is conducted via a simulator, we extended our 5-qubit MCZ experiment shown above to a noisy simulator. The results are shown in \cref{exp_5MC_noisy}. The proposed method still outperforms the full circuit execution even though the noise model provided by the manufacturer does not accurately capture the noise level exhibited by the real device.

\section{Conclusion}
\label{Conclusion}
In this work, we proposed an approach for cutting  multi-controlled Z gates by means of ZX-calculus based on the H-box fusion rule.
 We derived the upper bound $\kappa=6$ on the sampling overhead that is independent of the order of the gate. We validated the results on IBM hardware and found strong noise reduction due to the reduced amount of CNOT gates in the cut circuit. We anticipate the generalization of our method to multi-controlled rotation gates and extension to multi-qubit rotations. 
The optimality of the decompositions constructed in this work at present remains an open question.

\begin{acknowledgements}
We thank L.\ Burgholzer and  R.\ Wille for fruitful exchange and discussion.

We acknowledge the use of IBM Quantum services for this work. The views expressed are those of the authors, and do not reflect the official policy or position of IBM
or the IBM Quantum team.

The research was supported by the Bavarian Ministry of Economic Affairs, Regional Development
and Energy with funds from the Hightech Agenda Bayern and by the Bavarian Ministry for Economic Affairs,
Infrastructure, Transport and Technology through the
Center for Analytics-Data-Applications (ADA Center)
within the framework of “BAYERN DIGITAL II”.

The research was furthermore funded by the project QuaST, supported by the Federal Ministry for Economic Affairs and Climate Action on the basis of a decision
by the German Bundestag.
\end{acknowledgements}

\appendix

\section{Representation of an MCZ gate}
\label{Representation of a MCZ gate}
In this appendix, we prove the representation of an MCZ gate as stated in \cref{CZ-gate}.
We begin by introducing the diagrammatic notation of a general $n$-qubit column vector $\mathbf{v}$ as
\begin{equation}
\mathbf{v}=\sum_{i=0}^{2^n-1}v_i|i\rangle=\; \vcenter{\hbox{\includegraphics[width=0.87cm]{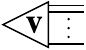}}}\;
= \;  \left(\vcenter{\hbox{\includegraphics[width=0.87cm]{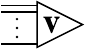}}}\right)^\dagger\,.
\end{equation}
With this definition and the $Z$-spider
\begin{equation}
 \vcenter{\hbox{\includegraphics[width=0.67cm]{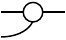}}}=\sum_{j\in\{0,1\}}|j\rangle\langle jj|\,,
 \label{copy_tensor}
\end{equation}
see \cref{Z_spider},
we obtain
\begin{align}
    \vcenter{\hbox{\includegraphics[width=1.43cm]{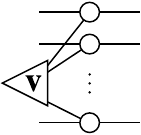}}}&=\Bigl(\sum_{j\in\{0,1\}}|j\rangle\langle jj|\Bigl)^{\otimes n}\sum_{i\in \{0,1\}^n} v_i \, \bigotimes_{k=1}^n\left(\mathbb{I}\otimes |i_k\rangle\right)\\
    &=\sum_i v_i |i\rangle\langle i|=\mathrm{diag}(\mathbf{v})\,.
    \label{diag}
\end{align}
To guarantee that the correct indices are contracted, we inserted one-qubit identities $\mathbb{I}$.
Consequently, \cref{diag} is proportional to a unitary matrix if all elements of $\mathbf{v}$ have the same absolute value or to a projector on a computational basis state if all but one element of $\mathbf{v}$ are zero.
We are now in the position to prove
\cref{CZ-gate} by setting
\begin{equation}
\label{Hbox_vector}
\vcenter{\hbox{\includegraphics[width=0.87cm]{figures/multi_qubit_vector_1.pdf}}}\;\;=\;\;\vcenter{\hbox{\includegraphics[width=0.7cm]{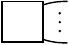}}}\;\;=\;\;(1,...,1,-1)^T
\end{equation}
where we recalled the definition of an H-box from \cref{Hbox}. Consequently, from \cref{diag}, we find $\mathrm{diag}(1,...,1,-1)$ for \cref{CZ-gate}, the matrix representation of an MCZ gate.

\section{Proof of H-box fusion rule}
\label{Proof of H-box fusion rule}
We proof \cref{fusion_rule} from the right to the left. We first define the joint indices $i=(i_1,...i_m)$ and $j=(j_1,...,j_n)$ as well as
$\pi_i=\Pi_{k=1}^m i_k$ and $\pi_j=\Pi_{k=1}^n j_k$.
We then find in Dirac notation with the indices $a,b,a^\prime$ and $b^\prime$  running over $\{0,1\}$, $j\in \{0,1\}^n$ and $i\in \{0,1\}^m$
\begin{align}
    &\sum_{i,j, a,b} (-1)^{\pi_i a}|i\rangle\langle a|\otimes (-1)^{\pi_j b}|j\rangle\langle b|\sum_{a^\prime,b^\prime}(-1)^{a^\prime b^\prime}| a^\prime b^\prime\rangle\\
    &\quad\quad=\sum_{i,j}\sum_{a,b}(-1)^{ab+\pi_ia+\pi_jb}| ij\rangle\\
    &\quad\quad=2\sum_{i,j}(-1)^{\pi_i\pi_j}| ij\rangle
\end{align}
which is equivalent to the left-hand side of \cref{fusion_rule}. In the last step we recalled that $\pi_i,\pi_j\in \{0,1\}$  so that 
\begin{align}
    &\sum_{a,b}(-1)^{ab+\pi_ia+\pi_jb}\\
    &\quad\quad=(-1)^{-\pi_i\pi_j}\sum_{a,b}(-1)^{(a+\pi_j)(b+\pi_i)}\\
    &\quad\quad=2(-1)^{\pi_i\pi_j}\,.
\end{align}
Alternatively, the H-box-fusion rule can be derived by the Schmidt decomposition of the vector $(1,...,-1)^T$ representing the diagonal elements of the unitary corresponding to the MCZ gate.

\section{H-box identities}
\label{H-box identities}
In this appendix, we prove \cref{main_identity1} and \cref{main_identity2}. The contraction of a single-qubit vector with an H-box is readily calculated as
\begin{align}
\label{Hbox_vector_contraction}
\vcenter{\hbox{\includegraphics[width=1.55cm]{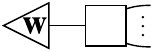}}}&=(w_0+w_1,...,w_0+w_1,w_0-w_1)^T\\
\intertext{by matrix-vector multiplication. Consequently, making use of \cref{diag}, we find}
\label{Hbox_contraction}
\vcenter{\hbox{\includegraphics[width=2.23cm]{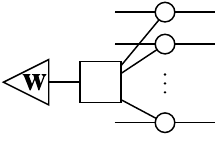}}}&=\mathrm{diag}(w_0+w_1,...,w_0+w_1, w_0-w_1)\,.
\end{align}
Therefore, \cref{Hbox_contraction} is proportional to a unitary for
$|w_0+w_1|=|w_0-w_1|$ and to a projector on the state $|1...1\rangle$ for $w_0=-w_1$.
In terms of spiders, these two conditions are either satisfied for 
$\mathbf{w}=\vcenter{\hbox{\includegraphics[width=0.6cm]{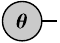}}}=\sqrt{2}\mathrm{e}^{i\theta/2}(\cos[\theta/2], -i\sin[\theta/2])^T$
or for $\mathbf{w}= \vcenter{\hbox{\includegraphics[width=0.6cm]{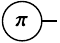}}}
=(1,-1)^T$. The former results in \cref{main_identity1} and the latter in \cref{main_identity2}.

\section{Rewriting the Projector}
\label{Rewriting the Projector}
To prove \cref{projector1}, we first restate the definition of $\mathcal{Z}$ from \cref{Definition_ZE}
\begin{equation}
\label{proof_ZE2}
    \mathcal{Z}(\rho)=\frac{1}{2^n}\sum_{k \in \{0,1\}^n}Z^{k_1}\otimes...\otimes Z^{k_n}\rho  Z^{k_1}\otimes...\otimes Z^{k_n}
\end{equation}
for an $n$-qubit state $\rho$. For a single-qubit state $\rho_1$ we note that
\begin{equation}
\label{proof_ZE1}
\sum_{k \in \{0,1\}}Z^{k}\rho_1  Z^{k} =2\sum_{j \in \{0,1\}}|j\rangle\langle j|\rho_1|j\rangle\langle j|
\end{equation}
 by substituting $Z=|0\rangle\langle 0|-|1\rangle\langle 1|$ as well as $\mathbb{I}=|0\rangle\langle 0|+|1\rangle\langle 1|$ into the left-hand side of  \cref{proof_ZE1}. Consequently, by applying \cref{proof_ZE1} iteratively, \cref{proof_ZE2} becomes
\begin{equation}
        \label{proof_ZE3}
   \mathcal{Z}(\rho) = \sum_{j\in\{0,1\}^n}|j\rangle\langle j|\rho |j\rangle\langle j|\,.
\end{equation}
Finally,  we rewrite \cref{proof_ZE3} as
\begin{align}
    \mathcal{Z}(\rho)=&2|1...1\rangle\langle 1...1|\rho |1..1\rangle\langle 1..1|\\
    &+ \sum_{j\in\{0,1\}^n}\beta_j|j\rangle\langle j|\rho |j\rangle\langle j|\\
    &=2\mathcal{P}_{1...1}(\rho)+\mathcal{P}(\rho)
\end{align}
where $\beta_j=-1$ for $j=(1,...,1)$ and $\beta_j=1$ otherwise, which proves \cref{projector1}.

\section{Derivation of sample-complexity overhead}
\label{Sampling-complexity overhead}
This appendix investigates in more detail the sampling overhead associated with evaluating a cut circuit, closely following Refs.\ \cite{Piveteau_2020, Peng2020}. In \cref{Sample complexity for circuit sampling} we  prove the statements on sampling overhead made in the main text for the case where in each experimental run a circuit from the decomposition is sampled from a probability distribution. Subsequently, in \cref{Sample complexity by pre-estimating circuits} we bound the variance by pre-estimating the expectation values of the partitioned circuits.

\subsection{Sample complexity for circuit sampling}
\label{Sample complexity for circuit sampling}
Consider an $n$-qubit quantum circuit. Assume the initial state $|0\rangle^{\otimes n}$ and consider the post-processing function $f:\{0,1\}^n \rightarrow [-1,1]$ on the measured bitstring $s$. Furthermore, assume that $f(s)$ can be efficiently calculated classically. These definitions describe the computational model of Refs.\ \cite{Bravyi2016, Peng2020}.
The post-processing function gives rise to an observable $O$ via
\begin{equation}
\label{definition post-processing}
     O=\sum_s f(s) P_s\,
\end{equation}
where $P_s$ is the projector on the computational basis state corresponding to bitstring $s$. Conversely, for instance, if the observable is a Pauli string, it can be written in the above form by local diagonalization and viewing the unitary diagonalization matrix
 as part of the circuit.
The goal of the quantum computation is to approximate $\langle O \rangle$ to additive error $\epsilon$ with high probability for which we will define statistical estimators in the following.

Cutting a gate amounts to replacing its superoperator by a decomposition as in \cref{general_decomposition_formula}.
We first consider the case where all $\mathcal{F}_i$ in the decomposition correspond to unitary gates. 
With this substitution, we find
\begin{equation}
\label{expectation value}
\langle O\rangle =\sum_i a_i \langle O\rangle_i=\sum_{i,s} p(s|i) a_i f(s)
\end{equation}
where $\langle . \rangle_i$ denotes the expectation value with respect to the state evolved by circuit $i$, containing $\mathcal{F}_i$. Moreover,
$p(s|i)$ is the probability to measure bitstring $s$ given circuit $i$.
To evaluate the result on a quantum device, we sample $i$ at each experimental run according to the probability distribution 
\begin{equation}
\label{probability for circuit}
p(i)=|a_i|/\kappa\,.
\end{equation}
We therefore define a random variable $I$ that takes values $i$ with probability $p(i)$ and the estimator \cite{Pashayan_2015}
\begin{equation}
\label{estimator1}
    \hat{f}=\kappa \, \mathrm{sign}(a_I) f(S_I)
\end{equation}
where the random variable $S_{I=i}$ models the bitstring outcomes of the circuit $i$.
This estimator is unbiased since
\begin{align}
\label{totalExpectation1}
   \mathbb{E}(\hat{f})&=\sum_{i,s} p(s,i)\kappa\,\mathrm{sign}(a_i) f(s)\\
   \label{totalExpectation2}
   &=\sum_{i,s}\kappa \,\mathrm{sign}(a_i)p(i)p(s|i)f(s)\\
   \label{totalExpectation3}
   &= \sum_{i,s}a_i p(s|i) f(s)\\
   \label{totalExpectation4}
   &=\langle O\rangle\,.
\end{align}
Here, we substituted \cref{probability for circuit} into \cref{totalExpectation2}. The final result is obtained by comparing \cref{totalExpectation3} to \cref{expectation value}.
The estimator for $N$ shots is given by the sample mean of \cref{estimator1}. Thus, to estimate $\langle O\rangle$ to additive error $\epsilon$ with probability $1-\delta$, Hoeffding's inequality provides the required number of experimental repetitions as
\begin{equation}
    N\geq 2\frac{\kappa^2}{\epsilon^2}\mathrm{ln}\left(\frac{2}{\delta}\right)
    \label{sample_overhead}
\end{equation}
where we used that $|\hat{f}|\leq \kappa$. The number of samples needed for the original circuit is obtained for $\kappa=1$ in \cref{sample_overhead} from which we infer the sampling overhead $ \mathcal{O}(\kappa^2)$.

It is straightforward to generalize this derivation to $K$ cuts. In this case, the expectation value of the observable is obtained as
\begin{equation}
\label{expectation value K cuts}
    \langle O \rangle = \sum_{i_1,...,i_K}\sum_s p(s|i_1,...,i_K)a_{i_1}\cdot ... \cdot a_{i_K}f(s)
\end{equation}
and the estimator for $K$ cuts becomes
\begin{equation}
\label{estimator K cuts}
    \hat{f}=\kappa^K \mathrm{sign}(a_{I_1})\cdot...\cdot \mathrm{sign}(a_{I_K})f(S_{I_1,...,I_K})
\end{equation}
where we defined the independent random variables $I_1$,...,$I_K$ that determine the specific circuit to run. To show equality between the expectation value of \cref{estimator K cuts} and \cref{expectation value K cuts}, we follow \cref{totalExpectation1} to \cref{totalExpectation4} and make use of the independence of $I_1$,...,$I_K$. Since $|\hat{f}|\leq \kappa^K$, Hoeffding's inequality applied to the sample mean of \cref{estimator K cuts},  provides the bound  $\mathcal{O}(\kappa^{2K})$ on the sampling overhead.
If the observable factorizes over the partitions $A$ and $B$ of the original circuit, the post-processing function factorizes as well, that is $f(s)=f^A(s^A)f^B(s^B)$ where $s^A$ and $s^B$ are the bitstring results on partition $A$ and $B$. If the initial state factorizes and all gates connecting the two partitions are cut, the partitioned circuits can be evaluated on independent quantum computers or sequentially on the same device.

We now turn to a quantum circuit that contains projectors.
Consider the map $\mathcal{M}$ consisting of a complete set of projectors $P_l$ and $\xi_l\in [-1,1]$ with
\begin{equation}
    \mathcal{M}(\rho)=\sum_l \xi_l P_l \rho P_l
    \label{measurment_channel}
\end{equation}
for state $\rho$. This map is neither positive since $\xi_l$ can be smaller than zero, nor trace preserving since the state in \cref{measurment_channel} is not normalized  after the projection and multiplied by $\xi_l$. Note that both $\mathcal{P}_{1...1}$ and $\mathcal{P}$ of \cref{projector1} are instances of $\mathcal{M}$. For $\mathcal{P}_{1...1}$  we set $\xi_l=1$ only for $l=(1,...,1)$ and zero otherwise. On the other hand for $\mathcal{P}$ we have $\xi_l=-1$ for $l=(1,...,1)$ and $\xi_l=1$ otherwise. 
Even though $\mathcal{M}$ does not correspond to a physical time evolution, we can nevertheless estimate
\begin{equation}
\label{nonphysical estimator}
    \chi =\mathrm{tr}(O\mathcal{U}\circ \mathcal{M}(\rho))
\end{equation}
by repeated use of a quantum computer. In \cref{nonphysical estimator}, the density matrix
 $\rho$ is the state right before the projectors and $\mathcal{U}$ is the unitary channel corresponding to the unitary evolution $U$ afterwards until the final measurement. 
To estimate $\chi$ on a quantum computer, we perform intermediate measurements, and sample according to the estimator
\begin{equation}
    \hat{f}=\xi_L f(S)
    \label{estimator M}
\end{equation}
where the random variable $S$ describes the bitstring outputs of the final measurements as before and the random variable $L$ models the outcomes of the intermediate measurements. 
According to \cref{estimator M}
we have to add $-f(S)$ in the sample mean for $\hat{f}$ if we found the all-one state in the intermediate measurement and $f(S)$ otherwise.
This estimator is unbiased since
\begin{align}
\label{expectation_C_1}
    \mathbb{E}(\hat{f})&=\sum_{s,l}p(s,l)\xi_l f(s)\\
    \label{expectation_C_2}
    &=\sum_{s,l}\xi_l\, \mathrm{tr}(P_s U P_l\rho P_l U^\dagger)f(s)\\
    \label{expectation_C_3}
    &=\sum_s\mathrm{tr}(P_s \mathcal{U}\circ \mathcal{M}(\rho))f(s)\\
    &=\chi\,.
\end{align}
The first line, \cref{expectation_C_1},
 is the definition of the expectation value and in \cref{expectation_C_2} we substituted $p(s,l)=\mathrm{tr}(P_s U P_l\rho P_l U^\dagger)$ \cite{Nielsen2010}. In the final step, we identified $O=\sum_s  f(s) P_s$. The number of samples required to determine $\chi$ to given additive error is again determined by Hoeffding's inequality applied to the sample mean of \cref{estimator M}.
Since $|\xi_l|\leq 1$ by definition, \cref{sample_overhead} remains unchanged when some of the circuits contain projectors.

\subsection{Sampling complexity by pre-estimating circuits}
\label{Sample complexity by pre-estimating circuits}
In  \cref{Experiments on IBMQ}, we show experimental results for a circuit that disintegrates into two independent partitions $A$ and $B$. After cutting the single gate that connects the two partitions, we have to evaluate
\begin{equation}
\label{sum over expected circuit pairs}
    \langle O\rangle
    =\sum_i a_i \langle O^A\rangle_i\langle O^B\rangle_i\,.
\end{equation}
Rather than sampling a circuit pair $i$ for each experimental run as discussed in \cref{Sample complexity for circuit sampling} we pre-estimate the expectation values $\langle O^A\rangle_i$ and $\langle O^B\rangle_i$ for all the circuits first and subsequently restore the result of the original circuit with the help of \cref{sum over expected circuit pairs}. In the following, it is shown that the standard deviation of \cref{sum over expected circuit pairs} can be bounded by $\epsilon$ for a total number of experimental runs
\begin{equation}
N\geq4\frac{\kappa^2}{\epsilon^2}
\label{error_bound_eq}
\end{equation}
where we allocate 
\begin{equation}
\label{samples for each circuit}
N_i=\frac{1}{2}\frac{|a_i|}{\kappa} N
\end{equation}
 samples to each circuit of circuit pair $i$ to determine the expectation values $\langle O^A\rangle_i$ and $\langle O^B\rangle_i$. Note that $2\sum_i N_i=N$.
Next, we define the estimator
\begin{equation}
\label{estimator_precalculation}
    \hat{f}=\sum_i a_i \hat{f}_i^A \hat{f}_i^B
\end{equation}
where $\mathbb{E}(\hat{f}_i^A)=\langle O^A\rangle $ and $\mathbb{E}(\hat{f}_i^B)=\langle O^B\rangle$. Since $\hat{f}_i^A$ and  $\hat{f}_i^B$ are independent, this estimator is unbiased. The variance of $\hat{f}$ is calculated as
\begin{align}
\label{variance estimator1}
&\mathrm{Var}(\hat{f}_i^A \hat{f}_i^B)\\
\label{variance estimator2}
    &\quad\quad=\mathrm{Var}(\hat{f}_i^A)\mathbb{E}[(\hat{f}_i^B)^2]+\mathrm{Var}(\hat{f}_i^B)[\mathbb{E}(\hat{f}_i^A)]^2\\
    \label{variance estimator3}
    &\quad\quad\leq \mathrm{Var}(\hat{f}_i^A)+\mathrm{Var}(\hat{f}_i^B)\\
    \label{variance estimator4}
    &\quad\quad\leq \frac{2}{N_i}
\end{align}
In \cref{variance estimator1} we first factorized the variance, valid for independent random variables. \cref{variance estimator3} makes use of the bounds $\mathbb{E}[(\hat{f}_i^B)^2] \leq1$ and $|\mathbb{E}(\hat{f}_i^A)|\leq1$ since $|f_i^A|\leq 1$ and $|f_i^B|\leq 1$. Finally, $\mathrm{Var}(\hat{f}_i)\leq1/N_i$ for $N_i$ independent samples.
Consequently,
\begin{align}
    \mathrm{Var}(\hat{f})&=\sum_i a_i^2\mathrm{Var}(\hat{f}_i^A \hat{f}_i^B)\\
    &\leq 2\sum_ia_i^2/N_i\\
    &\leq \epsilon^2
\end{align}
using \cref{estimator_precalculation}, \cref{variance estimator4} and \cref{samples for each circuit}.

\newpage
\onecolumngrid

\section*{Supplementary material}
In this supplemental material, we provide more details on our simulations and experiments on the ibmq\_ehningen system.

\subsection*{Random circuits}
\begin{figure}[H]
    \centering
    \includegraphics[width=\linewidth]{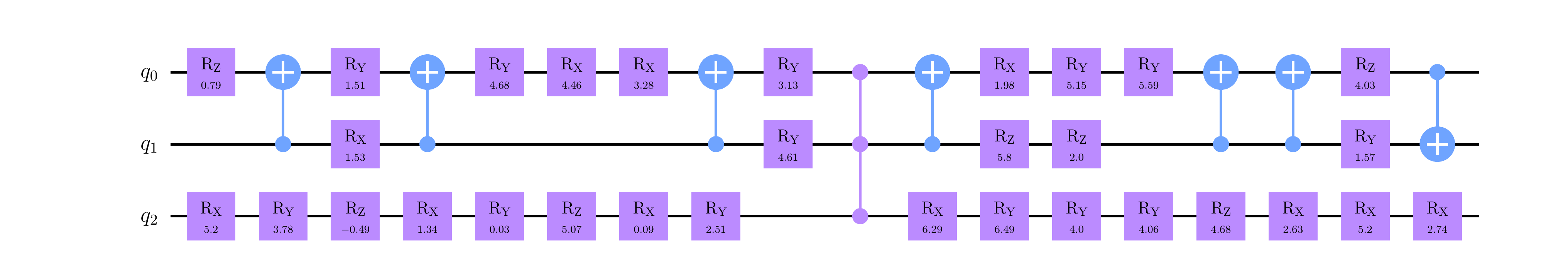}
    \caption{Random  three-qubit circuit with a CCZ gate at the center.}
    \label{circuit_CCZ_random}
\end{figure}

\begin{figure}[H]
    \centering
    \includegraphics[width=\linewidth]{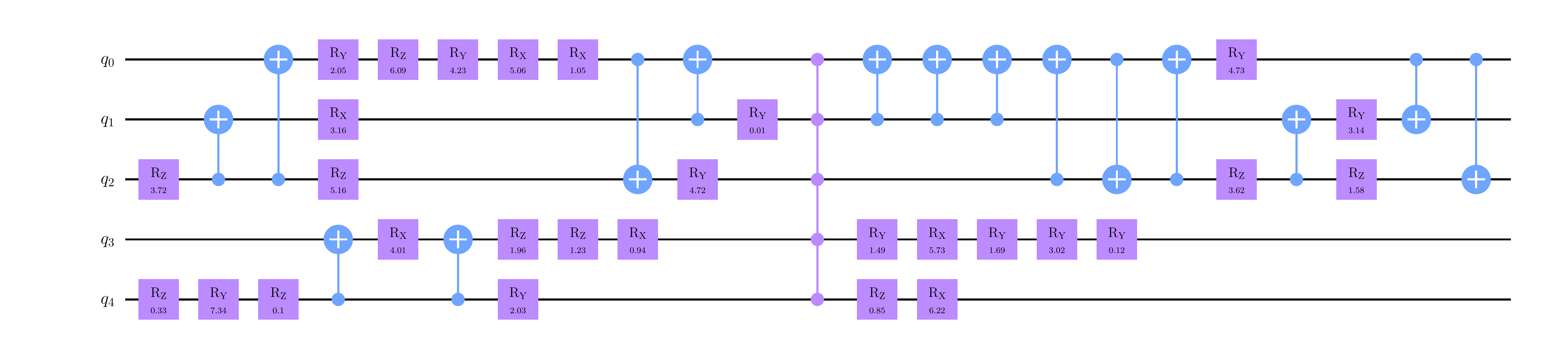}
    \caption{Random  five-qubit circuit with an MCZ gate of order five at the center.}
    \label{circuit_5MC}
\end{figure}

For our numerical simulations in Sec.~V of the main article we generated a set of random circuits consisting of about on average 30 single-qubit $\mathrm{R}_\mathrm{X}$, $\mathrm{R}_\mathrm{Y}$ and $\mathrm{R}_\mathrm{Z}$ rotations and 10 CNOT gates. We generated circuits with 3, 4, or 5 qubits with one MCZ gate in the center, accordingly, of order 3, 4, or 5. Two instances of these circuits are shown in \cref{circuit_CCZ_random} and \cref{circuit_5MC}. In the 3-qubit circuits,  one 1-qubit and one 2-qubit partition, and in the 5-qubit circuits one 2-qubit and one 3-qubit partition are only connected by the MCZ gate. For the numerical evaluation, we choose the Pauli $Z\otimes...\otimes Z$ string as observable. For generating the random circuits, we impose the constraint that the MCZ gate in the circuits should have a significant impact on the expectation value of the observable in order to be able to resolve its influence on the result with the available number of experimental runs. That is, the difference in the expectation value between a circuit with and without MCZ is chosen to be greater than 0.2. A proof-of-concept implementation of MCZ-gate cutting for arbitrary order of the gate is available on GitHub \cite{github}.

\subsection*{Experiments on ibmq\_ehningen}
\begin{figure}[H]
    \centering
    \includegraphics[width=\linewidth]{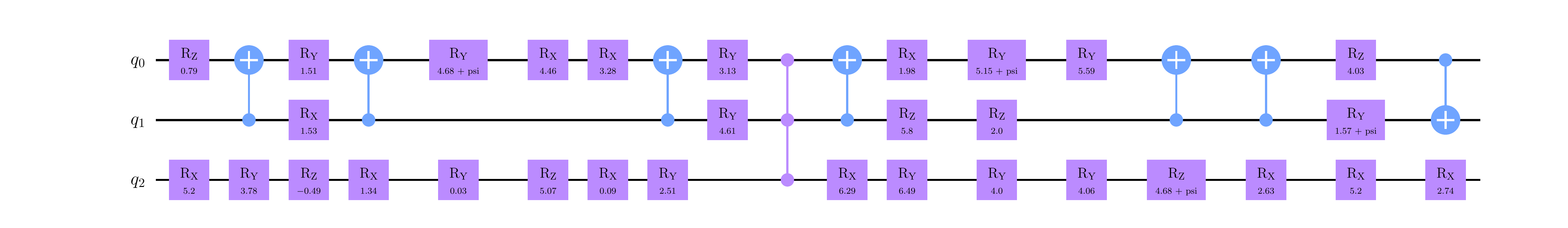}
    \caption{Circuit for the cutting of a CCZ gate. }
    \label{circuit_CCZ_scan}
\end{figure}

\begin{figure}[H]
    \centering
    \includegraphics[width=\linewidth]{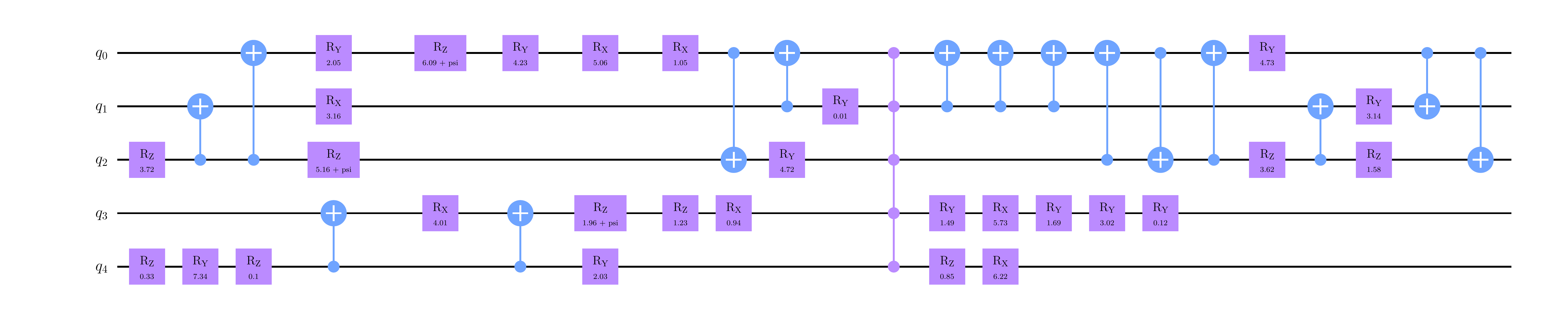}
    \caption{Circuit for the cutting of an MCZ gate of order 5.}
    \label{circuit_5MC_scan}
\end{figure}

\begin{figure}[H]
    \centering
    \includegraphics[width=8cm]{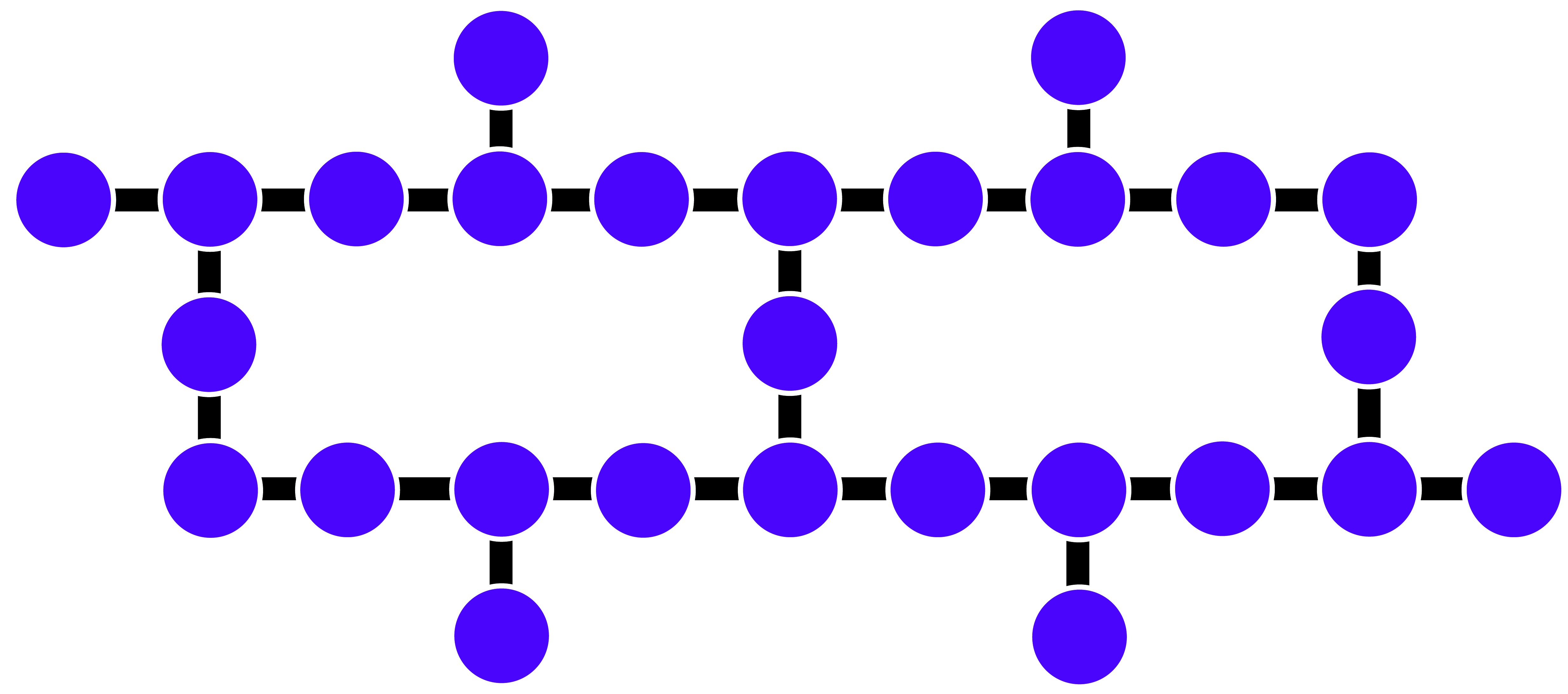}
    \caption{Connectivity graph of ibmq\_ehningen. }
    \label{connectivity}
\end{figure}

In this part of the supplemental material, we provide more details on the experimental setup. For the experiments we
    (1) again generated a random 3-qubit circuit and a 5-qubit circuit with two partitions, only connected by a CCZ gate and an MCZ gate of order 5, respectively. The circuits are shown in \cref{circuit_CCZ_scan} and \cref{circuit_5MC_scan}. Subsequently, we optimized the single-qubit rotation angles until the difference of the expectation values of a Pauli Z-string $Z\otimes...\otimes Z$ with respect to the circuit with and without the MCZ gate was maximal. Finally, we added one free parameter to the rotation angles of three single-qubit rotation gates, which were scanned in the experiments. The parameter is denoted by psi in the circuits shown in \cref{circuit_CCZ_scan} and \cref{circuit_5MC_scan}.
    (2) We used the highest level of optimization offered by the Qiskit API after transpiling the circuits to the native gate set and the hardware graph of the ibmq\_ehningen device.
    The highest level of optimization first searches for a layout that satisfies all the 2-qubit gate connectivity to the hardware graph considering the qubits readout errors and gate fidelities. Then the circuit is unrolled to the native gate set. Finally, optimizations in the form of commutative gate cancellation and re-synthesis of two-qubit unitary blocks are performed.
    (3) 18 data points between $[0, 2\pi]$ with equal intervals were chosen for scanning the free parameter.
    (4) We used $10^5$ shots per data point, the maximum number of shots per job allowed by the ibmq\_ehningen device.
    (5) All experiments were run with maximum error mitigation offered by ibmq wherever possible. We used the TREX option for readout-error mitigation \cite{vandenBerg2022} offered by Qiskit.
    (6)
    The ibmq\_ehningen device can be characterized by the following parameters \cite{ibmq2021}: It is a 27 qubit ibmq\_falcon processor, whose connectivity graph is shown in \cref{connectivity}. The noise parameters at the time of our experiments were the following:  Decoherence times: Average values $T_1=160$ \textmu s  and $T_2=150$ \textmu s with large fluctuations between the qubits. Single-qubit errors are of the order of $10^{- 4}$ and CNOT-gate errors of the order of $0.009$.

\end{document}